\documentclass[twocolumn,showpacs,superscriptaddress,preprintnumbers,amsmath,amssymb,prl]{revtex4}

\usepackage{graphicx}% Include figure files
\usepackage{dcolumn}% Align table columns on decimal point
\usepackage{bm}% bold math
\usepackage{multirow}
\usepackage{natbib}

\usepackage{color}
\usepackage[normalem]{ulem}

\newcommand{\gammadot}{\dot{\gamma}}
\newcommand{\sigmatot}{\sigma_{\rm tot}}
\newcommand{\sigmael}{\sigma_{\rm el}}
\newcommand{\sigmavisc}{\sigma_{\rm visc}}
\newcommand{\gammaeff}{\gamma_{\rm eff}}
\newcommand{\gammaload}{\gamma_{\rm y}}
\newcommand{\gammarex}{\gamma_{\rm dyn}}
\newcommand{\deltav}{|\Delta v|}

\begin{document}

%\preprint{/cond-mat.xxxxxxxx}

\title{Model for the Scaling of Stresses and Fluctuations in Flows near Jamming}

\author{Brian P. Tighe}
\affiliation{Instituut--Lorentz,
Universiteit Leiden, Postbus 9506, 2300 RA Leiden, The
Netherlands}
\author{Erik Woldhuis}
\affiliation{Instituut--Lorentz,
Universiteit Leiden, Postbus 9506, 2300 RA Leiden, The
Netherlands}
\author{Joris J.C. Remmers}
\affiliation{Instituut--Lorentz, Universiteit Leiden, Postbus
9506, 2300 RA Leiden, The Netherlands}
\affiliation{
%Department of Mechanical Engineering,
Eindhoven University of Technology,
Postbus 513, 5600 MB, Eindhoven, The Netherlands}
\author{Wim van Saarloos}
\affiliation{Instituut--Lorentz,
Universiteit Leiden, Postbus 9506, 2300 RA Leiden, The
Netherlands}
\author{Martin van Hecke}
\affiliation{
 Kamerlingh Onnes Lab, Universiteit Leiden, Postbus 9504, 2300 RA
Leiden, The Netherlands}

\date{\today}

\begin{abstract}
We probe flows of soft, viscous spheres near the jamming
point, which acts as a critical point for static soft spheres.
Starting from energy considerations, we find nontrivial scaling of
velocity fluctuations with strain rate. Combining this scaling with
insights from jamming, we arrive at an analytical model that
predicts four distinct regimes of flow, each characterized by
rational-valued scaling exponents. Both the number of regimes and
values of the exponents depart from prior results. We validate
predictions of the model with simulations.

\end{abstract}


\maketitle

The past few years have seen enormous progress towards
understanding the static, ``jammed'' state that occurs when soft
athermal particles are packed sufficiently densely that they
attain a finite rigidity
\cite{jamming,epitome,vanhecke10}.
Such systems may flow when shear stresses are applied,
and in seminal work, Olsson and Teitel addressed the relation between
strain rate, shear stress and packing fraction in a simplified
numerical model for the flow of soft viscous spheres \cite{olsson07}. When
rescaled appropriately, the data for strain rate $\gammadot$,
shear stress $\sigma$ and packing fraction $\phi$ were found to collapse
to two curves, reminiscent of second order-like scaling functions,
and a large length scale was found to emerge  near jamming.
Since then, qualitatively similar results have been obtained in
simulations of a number of flowing systems
\cite{haxton07,hatano08,heussinger09,otsuki09,ciamarra09},
but there is little agreement on the actual value of scaling
exponents, nor on the relation to jamming in static systems.

Here we describe an analytical model that connects the scaling of
static systems to the scaling of both the velocity fluctuations
and the shear stress of flowing systems near jamming. 
The model is built around a ``viscoplastic''
effective strain $\gammaeff = \gammaload +
\gammarex$, where $\gammarex$ is a dynamic contribution set by the
strain rate, and $\gammaload$ stems from the (dynamical) yield stress
and is controlled by the distance to jamming. We show that
steady state power balance dictates nontrivial scaling of
$\gammarex$ with strain rate, and propose a nonlinear
stress-strain relation that leads to a closed set of equations
predicting a rich scaling scenario for flows near jamming. 
We verify central ingredients
of the model and our predictions for the rheology
numerically in Durian's bubble model for foams 
\cite{durian95}. 
Our simple model captures and predicts the rheology and fluctuations
starting from the microscopic interactions; 
it also indicates the need for, and provides, new ways to present and
analyze rheological data near jamming.

{\em Numerical Model ---} The two-dimensional Durian bubble model
stipulates overdamped dynamics in which the sums of elastic and
dissipative forces on each bubble, represented by a disk,
balance at all times \cite{durian95}.
Forces are pairwise and occur only between
contacting bubbles. Elastic interaction forces are proportional to the disk
overlap, $f^{\rm el}_{ij} = k(R_i + R_j-r_{ij})^{\alpha_{\rm el}}$, where ${\vec
r}_{ij} := {\vec r}_j - {\vec r}_i$ points from one
bubble center to another and $R_i$ labels the radius of disk $i$.
In the full model that we focus on here,
viscous forces oppose the bubbles' relative velocity $\Delta {\vec
v}_{ij} := {\vec v}_j - {\vec v}_i$ with magnitude $f^{\rm
visc}_{ij} = b|\Delta v_{ij}|^{\alpha_{\rm visc}}$
\cite{footnote1}.
In simulations we set both exponents $\alpha_{\rm el}$ and 
$\alpha_{\rm visc}$ to unity. The strain rate $\gammadot$
is imposed via Lees-Edwards boundary conditions. The unit cell
contains a 50:50 bidisperse mixture of $N = 1020$ to 1210 bubbles with
size ratio 1.4:1. Stresses are averaged over a run 
(total time $20/\gammadot$) after discarding the transient.

\begin{figure}[tbp]
\centering
\includegraphics[clip,width=0.95\linewidth]{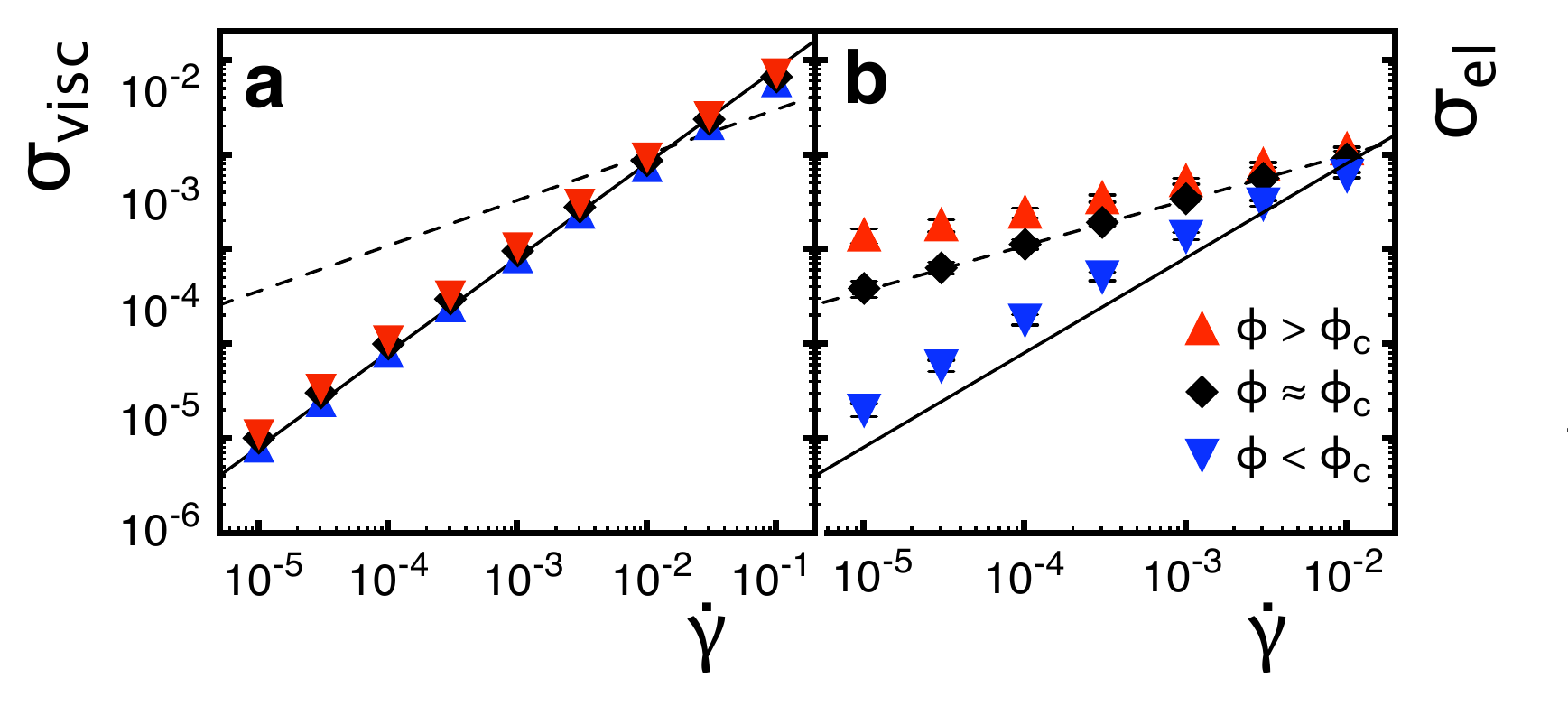}
\vspace{-0.45cm}
\caption{Viscous stress (a) and elastic stress (b)
for packing fractions $\phi = 0.80$, 0.8424, and 0.87. The dashed line
$\sim {\dot \gamma}^{0.48}$. $\sigmavisc$
scales linearly with $\gammadot$ (solid line)
and lies below the elastic stresses for
$\gammadot < {\cal O}(10^{-2})$.
} \label{fig:data}
\vspace{-0.45cm}
\end{figure}

{\em Elastic and Viscous Stress ---} Because forces balance on each bubble, the
shear stress can be computed according to $\sigma_{xy} := \sigmatot =
\frac{1}{2V}\sum_{\langle ij \rangle } r_{ij, x} \left( f^{\rm
el}_{ij, y} + f^{\rm visc}_{ij, y} \right)$, where $V$ is the area
of the unit cell and the sum runs over contacting pairs. It is
convenient to distinguish contributions of elastic and viscous
forces to the total stress $\sigmatot := \sigmael + \sigmavisc$.
Fig.~\ref{fig:data} depicts $\sigmael$ and $\sigmavisc$ as functions 
of strain rate for three
packing fractions. We find that the viscous stress $\sigmavisc$ depends only
weakly on $\phi$ and scales linearly with $\gammadot$, and
dominates the total stress for strain rates 
$\gammadot \gtrsim {\mathcal O}(10^{-2})$. We
denote this regime as Viscous (V), and since $\sigmavisc \sim
\gammadot$, $\sigmatot \sim \gammadot$ here. Our analytical model, developed below, treats the
case where $\sigmael$ dominates the stress, and
in all that follows, $\gammadot \le 10^{-2}$,
$\sigmavisc \ll \sigmael$, $\sigmatot \approx \sigmael$, and we no
longer distinguish between $\sigmatot$ and $\sigmael$, referring
to both as as $\sigma$. 

{\em Main Phenomenology ---} 
As Fig.~\ref{fig:data}b illustrates, 
the rheology is nontrivial and departs from simple linear scalings.
For $\phi$ above $\phi_c =  0.8423 \pm 0.001$
\cite{footnote2},
the stress flattens as $\gammadot$ is lowered, while for $\phi \approx
\phi_c$ the stress is well-described by a power law $\sigma \sim
\gammadot^{1/2}$ --- this will emerge below as the Critical
scaling regime of the rheology.  For $\phi$ below $\phi_c$,
the stress shows increasing downward curvature as strain rate is
decreased; our model does not treat this case, and we do not
consider it further.

\begingroup
\begin{table*}[t]
\begin{tabular*}{1.0\textwidth}{@{\extracolsep{\fill}} rcccc}

 & \bf Yield Stress & \bf Transition & \bf Critical & \bf Viscous
\\
\cline{2-5}

Stress &
\multicolumn{3}{c}{
\rule[2pt]{4cm}{0.1pt}\hspace{0.1cm}
$\sigma \approx \sigma_{\rm el} $
\hspace{0.1cm}\rule[2pt]{4cm}{0.1pt}} &
$\sigma \approx  \sigma_{\rm visc}$
 \\

\multicolumn{1}{c}{\multirow{2}{*}{Model}} &
\begin{minipage}{2.1cm}
\begin{flushleft}
\begin{equation}
\left \lbrace
\begin{array}{lcl}
\sigma \gammadot &\sim& \deltav^2 \nonumber \\
\gammaload &\sim& \Delta \phi \nonumber \\
\sigma &\sim& G \gammaload \nonumber
\end{array} \right.
\end{equation}
\end{flushleft}
\end{minipage} &

\begin{minipage}{2.6cm}
\begin{flushleft}
\begin{equation}
\left \lbrace
\begin{array}{lcl}
\sigma \gammadot &\sim& \deltav^2 \nonumber \\
\gammarex &\sim& \gammadot/\deltav \nonumber \\
\sigma &\sim& G \gammarex \nonumber
\end{array} \right.
\end{equation}
\end{flushleft}
\end{minipage} &

\begin{minipage}{2.5cm}
\begin{flushleft}
\begin{equation}
\left \lbrace
\begin{array}{lcl}
\sigma \gammadot &\sim& \deltav^2 \nonumber \\
\gammarex &\sim& \gammadot/\deltav \nonumber \\
\sigma &\sim& \gammarex^2 \nonumber
\end{array} \right.
\end{equation}
\end{flushleft}
\end{minipage} &

\\
\multicolumn{1}{c}{\multirow{2}{*}{Result}} &
\begin{minipage}{2.5cm}
\begin{flushleft}
\begin{equation}
\left \lbrace
\begin{array}{lcl}
\sigma  &\sim& \Delta \phi^{3/2}\nonumber \\
\deltav &\sim& \Delta \phi^{3/4}\gammadot^{1/2} \nonumber
\end{array} \right.
\end{equation}
\end{flushleft}
\end{minipage} &

\begin{minipage}{2.5cm}
\begin{flushleft}
\begin{equation}
\left \lbrace
\begin{array}{lcl}
\sigma  &\sim& \Delta \phi^{1/3} \gammadot^{1/3} \nonumber \\
\deltav &\sim& \Delta \phi^{1/6}\gammadot^{2/3} \nonumber
\end{array} \right.
\end{equation}
\end{flushleft}
\end{minipage} &

\begin{minipage}{2.5cm}
\begin{equation}
\left \lbrace
\begin{array}{lcl}
\sigma  &\sim& \gammadot^{1/2} \nonumber \\
\deltav &\sim& \gammadot^{3/4} \nonumber
\end{array} \right.
\end{equation}
\end{minipage} &

\begin{minipage}{2.5cm}
\begin{flushleft}
\begin{equation}
\left \lbrace
\begin{array}{lcl}
\sigma &\sim& \gammadot \nonumber \\
\deltav &\sim& \gammadot \nonumber
\end{array} \right.
\end{equation}
\end{flushleft}
\end{minipage}

\\ \noalign{\smallskip}

\multicolumn{1}{c}{\multirow{1}{*}{Range}} &
${\dot \gamma} <  \Delta \phi^\frac{7}{2}$ &
$\Delta \phi^\frac{7}{2} < {\dot \gamma} <  \Delta \phi^2$ &
$\Delta \phi^2 < {\dot \gamma} <  0.01$ &
$0.01 < {\dot \gamma}$
\\
\end{tabular*}
\vspace{-0.4cm}
\caption{Analytical model and its solutions in the Yield Stress, Transition,
Critical and Viscous regimes.}
\vspace{-0.5cm}
\label{tab:model}
\end{table*}
\endgroup

\begin{figure}[tbp]
\centering
\includegraphics[clip,width=0.95\linewidth]{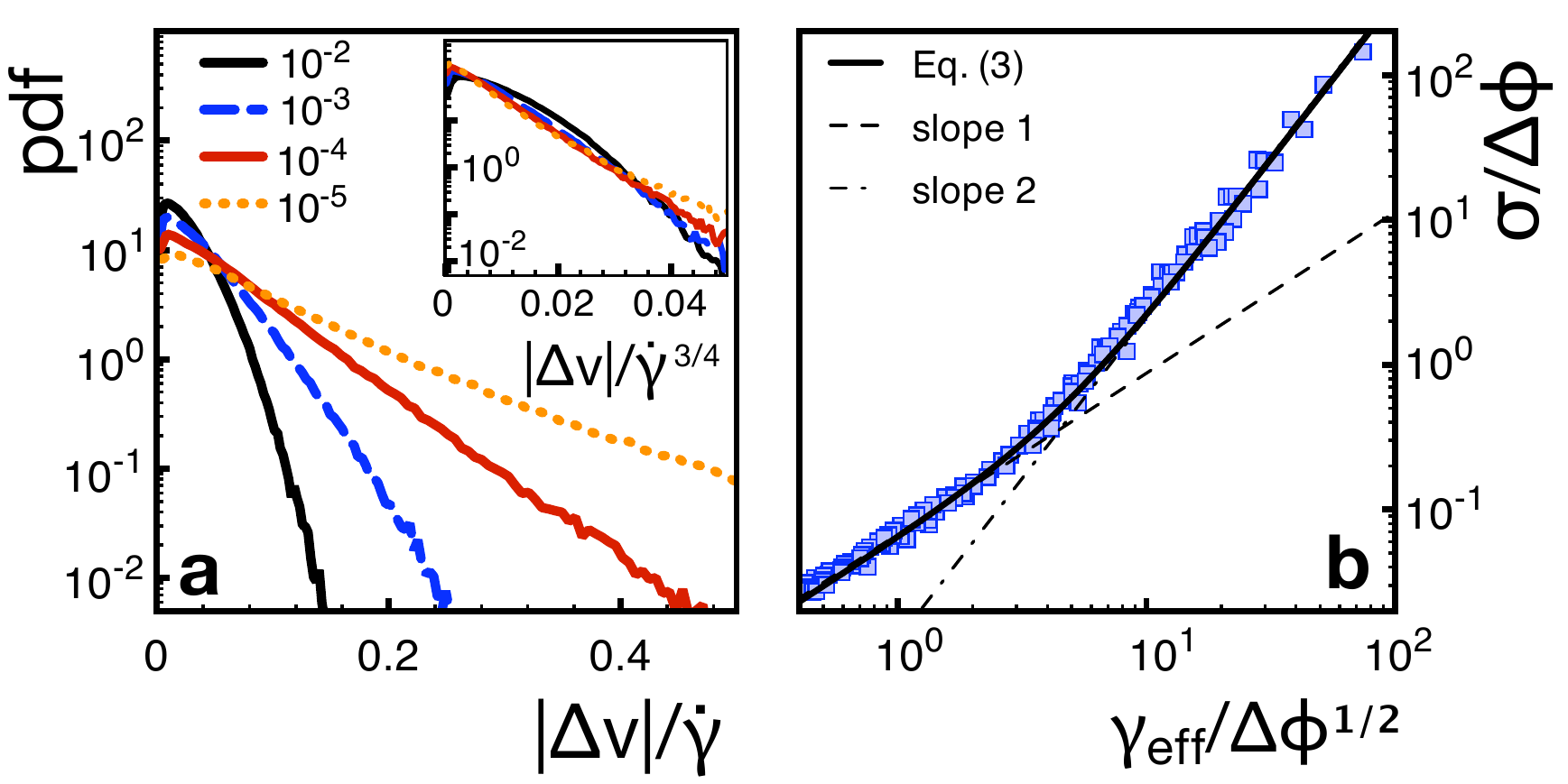}
\vspace{-0.3cm}
\caption{
(a) The probability
distribution function (pdf) of relative bubble velocities $\deltav$, 
for  $\phi \approx \phi_c$ and strain rates $\gammadot$ spanning
three decades (see legend), does not
collapse when rescaled by $\gammadot$. In contrast, the pdf of ${\deltav / \dot \gamma}^{3/4}$
yields a reasonable collapse (inset).
Note that Eq.~(\ref{balance}) predicts the pdf's second moment,
not its shape. (b) Collapsed stress-effective strain relation for 
$\gammaeff \propto \Delta \phi + 0.7 (\gammadot/\sigma)^{1/2}$ and 
the same data as in Fig.~\ref{collapse}. 
The solid curve is $y=0.085x \sqrt{1+0.05 x^2}$.
}
\label{newfig2}
\vspace{-0.45cm}
\end{figure}

{\em Analytical Model ---}
To relate stress and strain rate, we will {\em{(i)}} find expressions for the 
dynamic and yield strains $\gammaload$ and $\gammarex$ 
that constitute the effective strain and connect them to the stress 
via power balance; {\em{(ii)}} express the
stress as a function of $\gammaeff$.

{\em{(i)}} The effective strain $\gammarex$ can be thought of as
the typical strain undergone between plastic rearrangements of
the contact network. Because bubbles move
with relative velocity $\deltav$, the contact network rearranges
on a time scale $t_{\rm dyn} \sim {\bar d}/\deltav$,
where $\bar d$ is the average bubble diameter.
The typical strain incurred on this time
scale is $\gammarex := \gammadot t_{\rm dyn} \sim
\gammadot/\deltav$.

To obtain an estimate for $\deltav$,
we  turn our attention to the relation
between the scaling of stress and velocity fluctuations.
As noted in Ref.~\cite{ono03}, mechanical energy is supplied
to the system at a rate $\sim
\sigma_{\rm tot} \gammadot$. Energy dissipation
takes place by bubbles moving past each other --- hence the
dissipation rate scales as $ f^{\rm visc} \deltav
\sim  \deltav^2$ \cite{footnote3}.
Balancing the two yields
\begin{equation}\label{balance}
\sigma_{\rm tot}  \gammadot \sim  \deltav^2  \,.
\end{equation}
Eq.~(\ref{balance}) is the first of three relations comprising
our model:
given the stress, it provides the scaling of
$\deltav$. 

It will emerge from our model that the nontrivial scaling of velocity fluctuations 
underlies the rich rheology. Fig.~\ref{newfig2}a shows probability distributions
of $\deltav/\gammadot$ for $\phi \approx \phi_c$ and
$\gammadot \le 10^{-2}$. Since here $\sigma \sim \gammadot^{1/2}$,
$\deltav$ does {\em not} scale as $\gammadot$. In fact,
Eq.~(\ref{balance}) predicts that in this case $|\Delta v| \sim
\gammadot^{3/4}$. This gives a good collapse of the data
(inset Fig.~\ref{newfig2}a). Note that in the Viscous regime where
$\sigma \sim \gammadot$, one finds that $ \deltav^2
\sim \gammadot^2$, so that the typical relative velocity $\deltav$ scales 
trivially with $\gammadot$ (not shown). 

For $\phi > \phi_c$ one anticipates a threshold (yield) stress 
even for vanishingly low strain rate, which in our picture translates into an
additional contribution to the effective
strain; this is $\gammaload$. A reasonable expectation for the scaling of
$\gammaload$ is the strain scale required to prepare a packing at
$\phi = \phi_c + \Delta \phi$ by compressing a system from the
critical packing fraction:
$\gammaload \sim \Delta \phi /\phi \sim \Delta \phi$.
Collecting terms, the effective strain reads:
\begin{equation}
\gammaeff =  A_1 \Delta \phi + A_2 \bar{d}\, \gammadot /\deltav\,.  
\label{gdef}
\end{equation}

\begin{figure}[t]
\includegraphics[clip,width=0.925\linewidth]{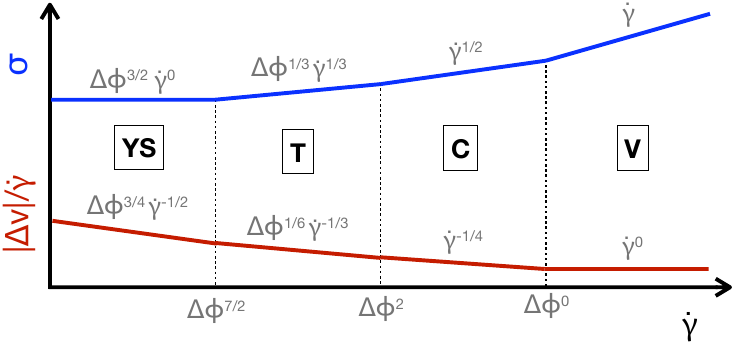}
\vspace{-0.3cm}
\caption{Schematic depiction of predictions
for scaling of stress $\sigma$ and
fluctuations in the relative bubble motion $|\Delta v|/\gammadot$
with strain rate $\gammadot$ and
distance to the critical packing fraction
$\Delta \phi = \phi - \phi_c$.
There are four distinct flow regimes: Yield Stress (YS),
Transition (T), Critical (C) and Viscous (V).
}
\label{fig:schematic}
\vspace{-0.45cm}
\end{figure}

{\em{(ii)}} We now construct a stress-strain relation 
$\sigma = g(\Delta \phi, \gammaeff) \, \gammaeff$
and make the ansatz that the shear modulus $g$ displays 
single parameter scaling: 
$g(\Delta \phi, \gammaeff) = \Delta \phi^p \,\tilde{g}(\gammaeff/\Delta \phi^q)$.
We will determine a form of $\tilde g$
based on known results for static systems.
Above $\phi_c$, static systems display a regime of linear response 
$\sigma = G\gamma$, where the static shear modulus 
$G = G_0 \sqrt{\Delta \phi}$ 
\cite{footnotesqrt}.
Hence $p=1/2$ and $\tilde{g}(x) \rightarrow G_0$ for $x \rightarrow 0$.
Precisely at $\phi_c$, $G$ vanishes, no analytic expansion of 
the stress-strain relation is possible, and 
critical static spring networks display the quadratic 
form $\sigma = \kappa |\gammaeff| \gammaeff$ \cite{wyart08}. 
It follows that $q = 1/2$ and $\tilde{g}(x) \rightarrow \kappa |x|$ 
for $x \rightarrow \infty$. 
Therefore the stress-strain relation can be rewritten as
\begin{equation}
\frac{\sigma}{\Delta \phi} =
\tilde{g} \left( \frac{\gammaeff}{\sqrt{\Delta \phi}} \right)
\frac{\gammaeff}{\sqrt{\Delta \phi}} \,.
\label{sdef}
\end{equation}

In the analysis to follow, only $p$, $q$, and the asymptotic scaling of 
$\tilde{g}(x)$ are essential. 
In Fig.~\ref{newfig2}b we show that a
scatter plot of $\sigma/\Delta \phi$ as function of
$\gammaeff/\sqrt{\Delta \phi}$ shows excellent data collapse with
the correct asymptotic behavior --- here 
$\gammaeff \propto \Delta \phi + (A_2 \bar{d}/A_1) (\gammadot /\sigma)^{1/2}$
and $A_2/A_1$ has been adjusted to obtain collapse.
We also note that the simplest choice for $\tilde{g}(x)$ 
that obeys reflection symmetry, remains
analytic above $\phi_c$ and obeys all necessary scalings is 
$\tilde{g}(x) =  G_0 \sqrt{1 + (\kappa x/G_0)^2 }$, which
fits the data remarkably well (Fig.~\ref{newfig2}b).

Our model comprises Eqs.~(\ref{balance}-\ref{sdef}), 
which express $\sigma$,
$\gammaeff$, and $\deltav$ in terms of $\gammadot$ and
$\Delta \phi$. For scaling analysis the constants
$A_1,A_2,G_0,\kappa$ and $\bar{d}$  can be
set to unity.

{\em Flow Regimes ---} The three equations for $\sigma$,
$\gammaeff$, and $\deltav$ in terms of $\gammadot$ and
$\Delta \phi$ lead to our scaling predictions.
Eqs.~(\ref{gdef}) and (\ref{sdef}) each have two scaling regimes, 
which are selected by varying $\gammadot$ and $\Delta \phi$.
In combination, these contribute three scaling regimes to 
the rheology, i.e.~the stress-strain rate relation. 
So where previous scaling ans\"atze presume two rheological regimes
\cite{olsson07,hatano08,otsuki09}
--- a yield stress plateau giving way to a power law in $\gammadot$ 
for higher strain rates --- we find Yield Stress
(YS), Transition (T), Critical (C) and Viscous (V) regimes, each
persisting over a finite range of strain rates (Fig.~\ref{fig:schematic}). 
Table I collects the pertinent equations, solutions
and parameter ranges for all scaling regimes.

\begin{figure}[t]
\centering
\includegraphics[clip,width=0.95\linewidth]{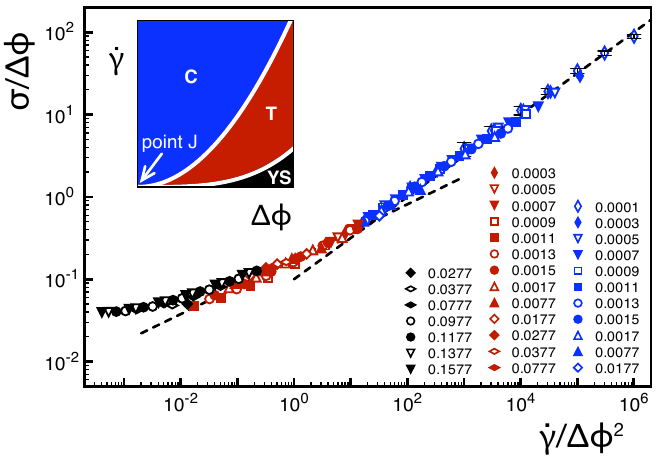}
\vspace{-0.3cm}
\caption{Scaling collapse over four decades in
strain rate $\dot \gamma$ and three decades in distance
to the critical packing fraction $\Delta \phi$ (legend). Rescaled
coordinates $\sigma/\Delta \phi$ and ${\dot \gamma}/\Delta \phi^2$
are appropriate for parameters spanning the Transition and Critical
regimes (see Table \ref{tab:model}). Dashed lines are guides to
the eye with slopes $1/3$ and $1/2$. (inset) Boundaries
between the Yield Stress, Transition and Critical regimes in the
$\Delta \phi$-$\gammadot$ plane. }
\label{collapse}
\vspace{-0.45cm}
\end{figure}

{\em Checking the Model ---} For data restricted to two
regimes, it is possible to collapse the $\gammadot$-$\sigma$ flow curves
to a master curve by rescaling with $\Delta \phi$.
For the Transition and Critical regimes, the rheology
is predicted to obey $\sigma \sim \Delta \phi^\frac{1}{3}\gammadot^\frac{1}{3}$ and
$\sigma \sim \gammadot^\frac{1}{2}$, respectively.
Hence
{\em for data in these two regimes},
$\sigma/\Delta \phi$ vs $\gammadot/\Delta \phi^2$ 
can be collapsed to a master flow curve,
characterized by a crossover from a $1/3$ to a $1/2$ power law scaling.

This strong test of the model is shown in Fig.~\ref{collapse},
where data for three decades in $\Delta \phi$ and four
decades in $\gammadot$ collapse to a single master
curve.
Data points satisfying $\gammadot < c_1\Delta \phi^{7/2}$ and
$\gammadot > c_2\Delta \phi^{2}$ \cite{footnote4}
are labeled Yield Stress (black)
and Critical (blue), consistent with scaling predictions; 
the Transition regime (red) lies in between.  YS data points
``peel off'' from the master curve
(note the black data points above red ones)
because it is not possible to collapse {\em three} regimes 
in one plot when their crossovers scale differently. 

We stress that the scaling exponents, including dependence on $\Delta \phi$,
are all predictions, not adjustable parameters. The excellent data
collapse in Figs.~\ref{newfig2}b and \ref{collapse} is therefore a striking confirmation
of the model.

We can gain some intuition for the various regimes
by considering different approaches to the critical point
(see Fig.~\ref{collapse} inset).
Fixing $\phi = \phi_c$ and adiabatically lowering the strain rate
approaches point J from the Critical regime
$\sigma \sim \gammarex^2 \sim \gammadot^{1/2}$,
where stress is always dominated by dynamic effects.
Similarly, fixing $\gammadot = 0^+$ and adiabatically decreasing
$\Delta \phi$ approaches point J from the Yield Stress regime
$\sigma \sim G\gammaload \sim \Delta \phi^{3/2}$, where the flow is
rate independent.
Finally, there is an anomalous flow regime
$\sigma \sim G\gammarex \sim \Delta \phi^{1/3} \gammadot^{1/3}$
that transitions between the Critical and Yield Stress regimes.
It is traversed when varying $\gammadot$ at finite $\Delta \phi$
or vice versa.

\begin{figure}[tbp]
\centering
\includegraphics[clip,width=0.925\linewidth]{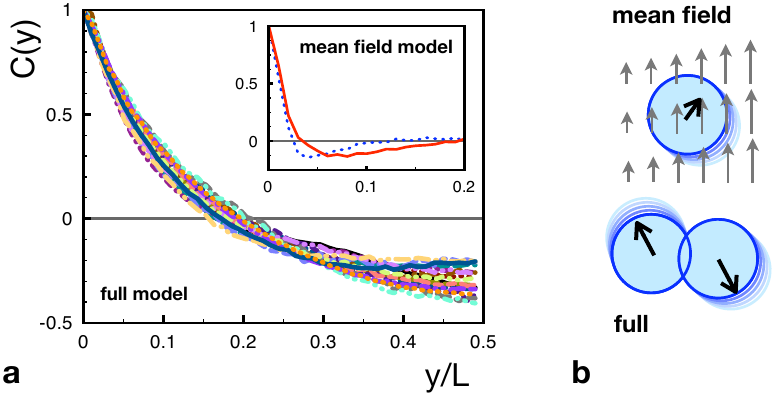}
\vspace{-0.3cm}
\caption{ (a) Two-point
correlation function $C(y)$ for $L = 75$
(as in Figs.~\ref{fig:data} and \ref{collapse}),
$\gammadot = 10^{-3}$ and $10^{-5}$, and $\phi = 0.82$, $\phi \approx \phi_c$
and $\phi = 0.86$. Also plotted are system sizes $L = 54$ and $105$ for
strain rates $\gammadot = 10^{-3}$ and $10^{-4}$ and packing fractions
$\phi = 0.82$, $0.84$, and $0.86$.
(inset) $C(y)$ has a different form in the mean field
model and evolves with $\gammadot$ and $\Delta \phi$. Here, two examples:
$\phi = 0.841$ (solid) and $0.88$ (dashed) at $\gammadot = 10^{-3}$.
(b) Dissipation in the models.
Viscous forces (black arrows) are proportional to the difference
between a bubble's own velocity and a linear velocity profile
(mean field model) or the velocity of a second bubble (full model).
}
\label{outlook}
\vspace{-0.45cm}
\end{figure}

{\em Length scale ---} In our model there is no
strain rate dependent length scale \cite{olsson07,lemaitre09}
to capture ``swirls'' or avalanching rearrangements.
To test this assumption, in Fig.~\ref{outlook}a we plot correlations
in the non-affine component of the bubble
velocities, $C(y) := \langle v_x(0) v_x(y) \rangle$. We find $C(y)$
to have a form that is ($i$) reminiscent of disordered static
\cite{didonna05} and quasistatic \cite{heussinger09}
soft sphere systems and ($ii$) remarkably robust to changes in
$\gammadot$ and $\phi$. Moreover, as in static linear response
\cite{didonna05},
$C(y/L)$ collapses for different box sizes $L$,
suggesting that for the system sizes studied here the box size is the
only relevant macroscopic length scale. We note that $C(y)$ was
measured in Ref.~\cite{olsson07} in the mean field (MF)
bubble model, which replaces realistic bubble-bubble viscous
forces by an effective drag term that punishes deviations from
an affine (linear) velocity profile --- see Fig.~\ref{outlook}b.
The minimum of $C(y)$ in the MF model (Fig.~\ref{outlook}a inset) selects a
length that was found to scale with $\Delta \phi$ and $\gammadot$ \cite{olsson07}.
Because this behavior vanishes when the mean field approximation is
lifted, we conclude that it is an artifact of the MF dynamics.

{\em Discussion and Outlook ---}
The unanticipated presence of {\em three}
rate dependent regimes with distinct rational-valued scaling exponents
offers an explanation for dissimilar exponents found in the literature
\cite{olsson07,haxton07,hatano08,otsuki09,xu05,durian95,lemaitre09}.
For data spanning some combination of the Transition,
Critical and Viscous regimes, one could fit a power law $\sigma \sim \gammadot^\beta$ with
effective exponent $1/3<\beta<1$. Reported scalings indeed range from around
$\beta \approx 0.4$ \cite{olsson07} to $0.6$ \cite{xu05,hatano08}
and even $1.0$ \cite{durian95}. 
Though the prediction is more difficult to test,
the model also provides a plausible argument that the dynamic yield
stress of systems with spring-like elastic interactions
has super-linear dependence on distance to $\phi_c$
($\sigma_{\rm y} \sim \Delta \phi^{3/2}$),
rather than linear \cite{lois07,otsuki09}. Refs.~\cite{olsson07,hatano08}
indeed find super-linear scaling, while $\sigma_{\rm y} \sim \Delta \phi$ in
the quasistatic simulations of Ref.~\cite{heussinger09}. 
Recent work by Hatano probing the lowest shear rates to date finds 
$\sigma_{\rm y} \sim \Delta \phi^\Delta$ with 
$\Delta = 1.5 \pm 0.1$ \cite{hatano10}.

The model is easily generalized to other microscopic interactions. Elastic
interactions enter through the scaling of the shear modulus 
$G \sim \Delta \phi^{\alpha_{\rm el}-1/2}$,
while different viscous force laws affect the fluctuations
via power balance: $\sigma \gammadot \sim \deltav^{\alpha_{\rm visc} + 1}$. 
Recent data for viscous NIPA particles are consistent both with our 
prediction of $\sigma \sim \gammadot^{1/2}$ in the Critical regime 
and with a yield stress $\sigma_{\rm y} \sim \Delta \phi^{\alpha_{\rm el} + 1/2}$ 
for Hertzian interactions, $\alpha_{\rm el} = 3/2$ \cite{nordstrom10}.
For physical foams, believed to have a viscous exponent
$\alpha_{\rm visc} = 2/3$ \cite{katgert08}, the
Critical regime scales as
$\sigma \sim \gammadot^{2\alpha_{\rm visc}/(\alpha_{\rm visc}+3)} \sim \gammadot^{4/11}$,
in remarkable agreement with recent experiments
that found $\sigma \sim \gammadot^{0.36}$ \cite{katgert08}. Finally,
for slow frictional flows, both the
the drag forces and the global rheology are rate independent 
\cite{schall10}. We suggest that the
global rate independence is not a triviality and note that it
is consistent with our
model, where $\sigma \sim \gammadot^0$ for $\alpha_{\rm visc} \rightarrow 0$.

\begin{acknowledgments}
We thank O.~Dauchot, J.M.J.~van Leeuwen, A.J.~Liu, T.C.~Lubensky,
M.E. M\"obius, S.R.~Nagel, P.~Olsson, S.~Teitel and Z. Zeravcic
for helpful interactions. Financial and computational
support from the Dutch physics foundation FOM, the Netherlands 
Organization for Scientific Research, and the 
National Computing Facilities Foundation are gratefully 
acknowledged.
\end{acknowledgments}
\vspace{-0.45cm}
\bibliographystyle{apsrev}
\bibliography{rheology}

\begin{thebibliography}{21}
\bibitem{footnote1}
\bibinfo{author}{ In simulations we fix $k = 0.35$, $b = 0.025$. Results are presented in dimensionless units where the smaller disk radius $R_s$ defines the unit of length, $k$ the unit of stress, and $b/k$ the unit of time. }
\bibitem{footnote2}
\bibinfo{author}{ We determine $\phi_c$ and its error bar from the range over which 
Fig.~\ref{collapse} shows good collapse.}
\bibitem{footnotesqrt}
\bibinfo{author}{
The square root scaling of $G$ with $\Delta
\phi$ (for spring-like interactions) 
is a hallmark of the static unjamming transition in both 
isotropic \cite{epitome,vanhecke10} and sheared packings \cite{ciamarra09}. }
\end{thebibliography}

\end{document}